\input{aipcheck}

\documentclass[final]{aipproc}

\layoutstyle{8x11double}

\begin{document}

\title{Simulations of Bunch Merging in a Beta Beam Decay Ring}


\author{D. C. Heinrich}{
  address={CERN, Geneva, Switzerland}
}
\author{C. Hansen}{
  address={CERN, Geneva, Switzerland}
}
\author{A. Chanc\'e}{
  address={CEA, IRFU, SACM, F-91191 Gif-Sur-Yvette, France}
}


\begin{abstract}
To further study neutrino oscillation properties a Beta Beam facility has been proposed. Beta decaying ions with high kinetic energy are stored in a storage ring ("Decay Ring") with straight sections to create pure focused (anti) electron neutrino beams. However to reach high sensitivity to neutrino oscillation parameters in the experiment the bunched beam intensity and duty cycle in the DR have to be optimized. The first CERN-based scenario, using $^{6}$He and $^{18}$Ne as neutrino sources, has been studied using a bunch merging RF scheme. Two RF cavities at different frequencies are used to capture newly injected bunches and then merge them into the stored bunches. It was shown that this scheme could satisfy the requirements on intensity and duty cycle set by the experiment. This merging scheme has now been revised with new simulation software providing new results for $^{6}$He and $^{18}$Ne. Furthermore bunch merging has been studied for the second CERN-based scenario using $^{8}$Li and $^{8}$B. 
\end{abstract}

\maketitle

\section{Introduction}

Improving the measurements of the known neutrino oscillation parameters 
(like $\Delta m_{21}^{2}$, $\left\vert \Delta m_{32}^{2}\right\vert$, 
$\theta_{12}$ and $\theta_{23}$) and determination of
the unknown ones (like $\theta_{13}$, sign$\left(\Delta m_{32}^{2}\right)$ and $\delta_{cp}$) 
requires precision measurements on a highly pure
and intense neutrino beam whose characteristics are well known. One
of the proposed next generation neutrino oscillation facilities \cite{FP7-EURONU}
is the Beta Beam concept \cite{ZUCCHELLI-BETABEAM}. Beta decaying
ions are stored at $\gamma=100$ in a horse-racetrack shaped storage
ring, the \textquotedbl{}Decay Ring\textquotedbl{}. One of the straight
sections is aimed at an oscillation experiment and the decaying ions
create a highly pure (anti) electron neutrino beam with an opening
angle of $1/\gamma$.

Within the EURISOL design study (FP6 \cite{FP6-HOMEPAGE})
the feasibility of a scenario using $^{6}$He
(as antineutrino source)\ and $^{18}$Ne (as neutrino source) isotopes
has been investigated. A 440kt \v{C}erenkov detector located in the
Fréjus tunnel at $L_{FP6}=130$~km distance from CERN was foreseen
to detect the incoming neutrinos. The proposed method for bunch injection
in the Decay Ring (DR), the so called \textquotedbl{}RF bunch merging
scheme\textquotedbl{}, creates bunches with sufficient
intensity while keeping the duty cycle (which coincides with the suppression
factor of the experiment) of the DR\ at $0.58\%$ and well below
the sensitivity threshold of 1\% given within FP6. 

The EUROnu design study (FP7 \cite{FP7-HOMEPAGE})
includes an additional scenario using
$^{8}$B (as neutrino source) and $^{8}$Li (as antineutrino source) isotopes. 
Since these isotopes
feature a higher Q value, the neutrino energies will be higher and
a longer baseline between production and detection is needed. The
detector can be placed in Gran Sasso ($L_{FP7}=732$~km) or Canfranc
($L_{FP7}=630$~km). A new RF scheme called \textquotedbl{}barrier
buckets\textquotedbl{} was proposed but simulations showed that an
optimization between bunched intensity and duty cycle could not be
achieved \cite{HANSEN}. This report therefore focuses on the adaptation
of the RF bunch merging scheme used in the EURISOL design study to
the isotopes used within EUROnu 

Both baselines assume the Beta Beam accelerator complex to be located 
at CERN, reusing the existing PS and SPS, and that the DR has the same circumference 
as SPS, namely 6911 m.  

\begin{center}
\begin{figure}[h]
\vspace*{-0.2em}
\begin{centering}
\includegraphics[width=0.46\textwidth]{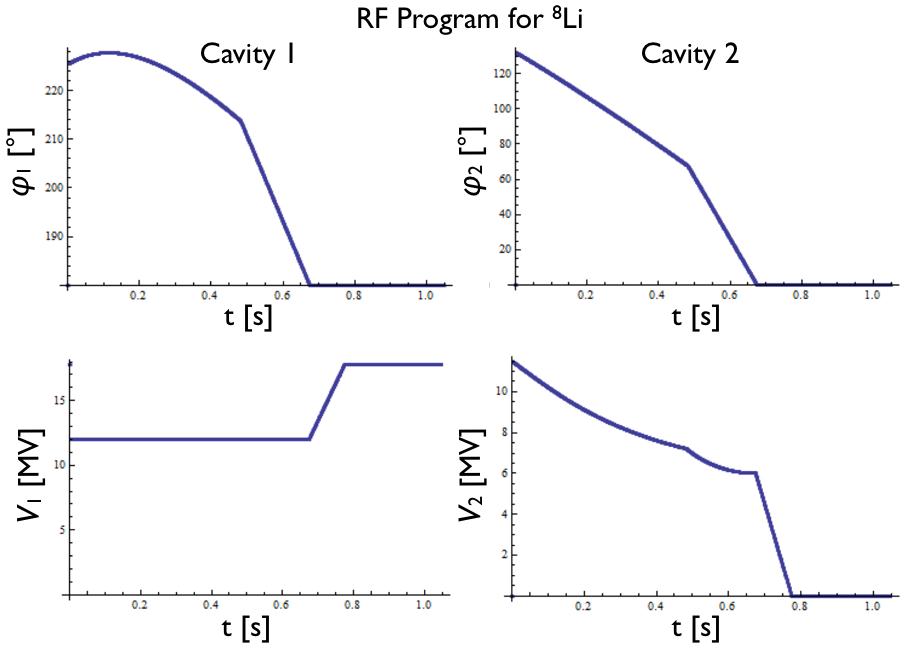}
\par\end{centering}
\caption{\label{fig:RF}Lithium RF program in terms of phases and voltages
for both RF systems as used in BBPhase.}
\end{figure}
\par\end{center}

\section{RF Simulations}

The RF simulations are done by using the 2D longitudinal phase space
program $\mbox{"BBPhase"}$ \cite{SVN} 
written to investigate the
possibility to use Barrier Buckets in the Decay Ring \cite{HANSEN}.
It has since been adapted to perform Bunch Merging simulations. 
The program tracks the particles on a turn-by-turn basis.
After each turn the particles' phases and momenta are evaluated dependent
on the voltage they have seen from the RF system in the previous turn.
The combined voltage from two cavities with frequencies $f_{1}=40$~MHz ($h_{1}=924$)
and $f_{2}=80$~MHz ($h_{2}=2h_{1}=1848$), is  
\begin{equation}
  V_{rf}=\sum_{i=1}^{2}V_{i}\sin\left(h_{i}\varphi+h_{i}\varphi_{i}\right)
\label{eq:volt}
\end{equation}
where the (time dependent) maximum voltages are $V_{i}$,
the RF phase when a reference particle passes through the long wavelength RF cavity is $-h_{1}\varphi_{1}$
and $-h_{2}\varphi_{2}$ when it passes through the fast frequency RF cavity
and $\varphi$ is the azimuthal difference between the reference particle and a given particle in the bunch.

The merging program was theoretically optimized for  $^{6}$He and $^{18}$Ne \cite{CHANCE} 
and has now been adapted for $^{8}$Li and $^{8}$B. 
The plots in fig. \ref{fig:RF} represent the changes in
phases and voltages for the merging program of Lithium.

\begin{figure}[ht]
$\begin{array}{c@{\hspace{0.1in}}c}
\includegraphics[angle=0, scale= .15]{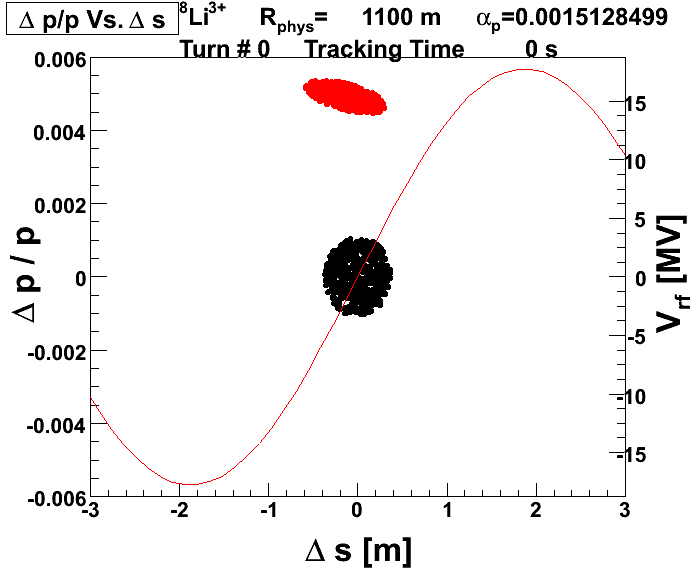}   &  \includegraphics[angle=0, scale= .15]{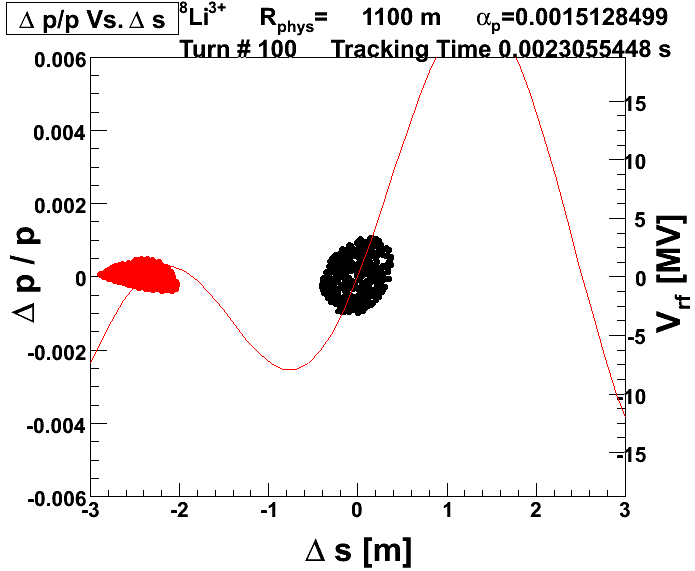} \\ [-2.7cm]
\multicolumn{1}{l}{\mbox{\bf \ \  \ \ \ \ \ \  (a)}}                                &  \multicolumn{1}{l}{\mbox{\bf \ \  \ \ \ \ \ \   (b)}} \\ [+2.3cm]
\includegraphics[angle=0, scale= .15]{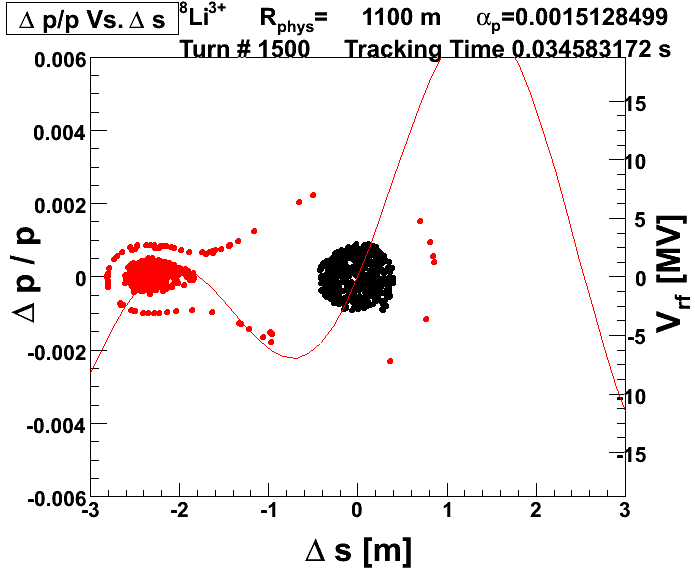}   &  \includegraphics[angle=0, scale= .15]{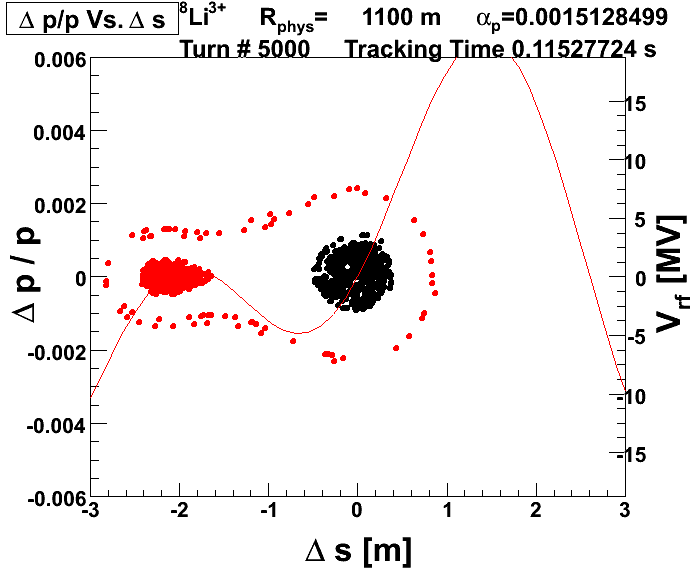} \\ [-2.7cm]
\multicolumn{1}{l}{\mbox{\bf \ \  \ \ \ \ \ \  (c)}}                                &  \multicolumn{1}{l}{\mbox{\bf \ \  \ \ \ \ \ \   (d)}} \\ [+2.3cm]
\includegraphics[angle=0, scale= .15]{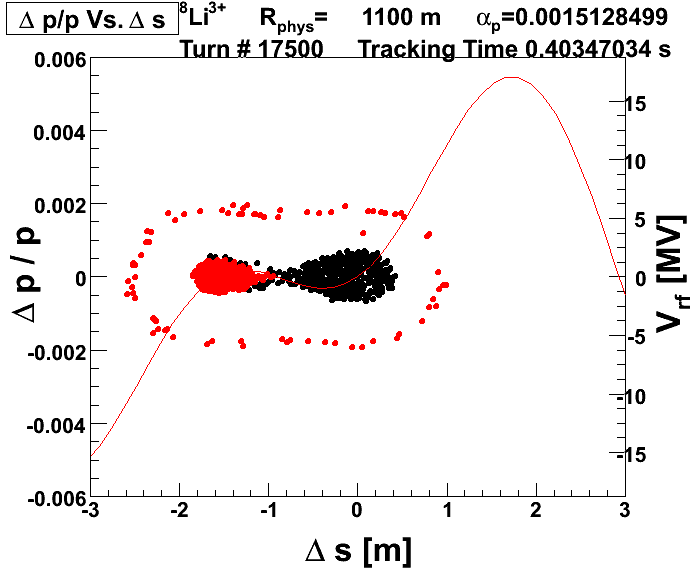}   &  \includegraphics[angle=0, scale= .15]{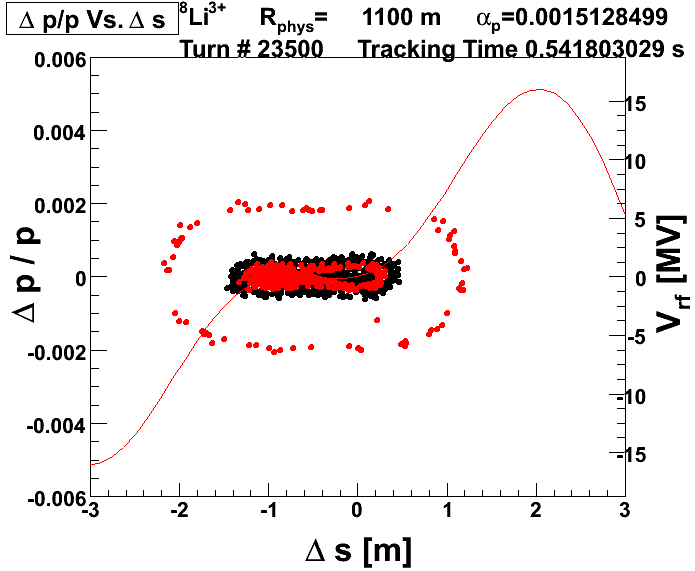} \\ [-2.7cm]
\multicolumn{1}{l}{\mbox{\bf \ \  \ \ \ \ \ \  (e)}}                                &  \multicolumn{1}{l}{\mbox{\bf \ \  \ \ \ \ \ \   (f)}} \\ [+2.3cm]
\includegraphics[angle=0, scale= .15]{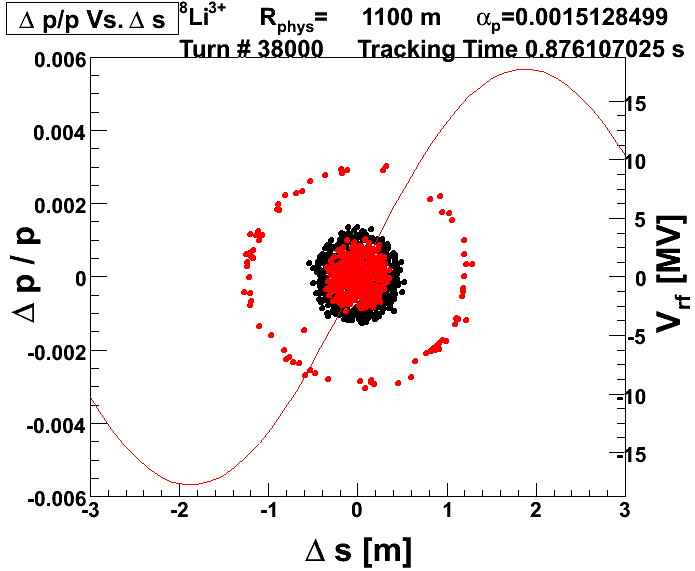}   &  \includegraphics[angle=0, scale= .15]{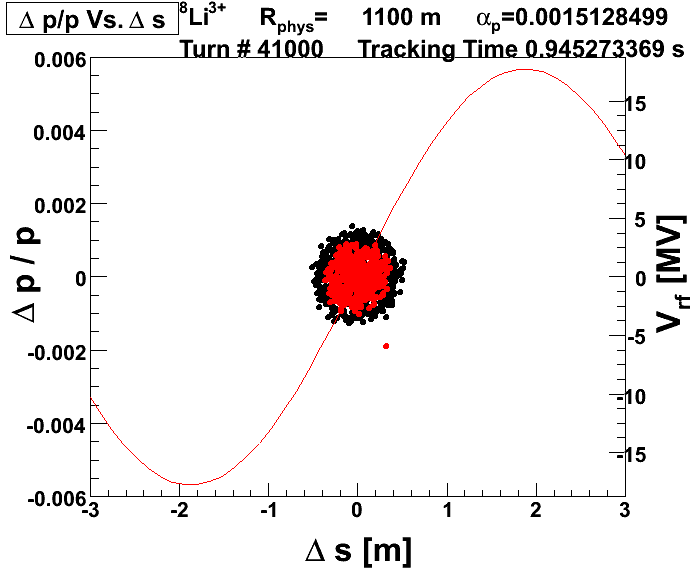} \\ [-2.7cm]
\multicolumn{1}{l}{\mbox{\bf \ \  \ \ \ \ \ \  (g)}}                                &  \multicolumn{1}{l}{\mbox{\bf \ \  \ \ \ \ \ \   (h)}} \\ [+2.3cm]
\end{array}$
\caption{Different phases of the RF Bunch Merging procedure.
The incoming bunch is injected (a) and then captured by the switched-on
second cavity after a quarter turn (b). Asymmetric (c, d, e) and symmetric
(f, g) merging are applied and the beam is collimated before the end
of the procedure (h).}
\label{fig:LI-MERGING}
\end{figure}

We apply the merging program (fig. \ref{fig:RF}) of Lithium to the
case that one bunch is already injected and circulating in the main
bucket (black dots in fig. \ref{fig:LI-MERGING} represent multiparticles
of this bunch). The newly injected beam (red) is distributed in a
longitudinal phase space ellipse at a momentum excess of $5$ per
mil (fig. \ref{fig:LI-MERGING}a). 
The best capture efficiency could be obtained by slightly lowering the particles
momentum excess to $4.92$ per mil and moving it about $9.6$ cm
behind the stored bunch. The particles perform a quarter synchrotron
turn and loose their elliptical distribution since the movement around
the synchronous particle is not linear in this region of phase space
(fig. \ref{fig:LI-MERGING}b). This causes some particles to be
outside the elliptical shaped capture bucket when the second cavity
is turned on. Those particles are not captured and circle
on the separatrix around both buckets. Asymmetric merging is applied
during 0.5 seconds where the main bucket shrinks
in size (until both buckets have the same size, fig. \ref{fig:LI-MERGING}c, \ref{fig:LI-MERGING}d and \ref{fig:LI-MERGING}e).
After that the symmetric merging procedure takes about $0.2$ seconds
to decrease the distance between the two buckets to zero (fig. \ref{fig:LI-MERGING}f and \ref{fig:LI-MERGING}g).
At this point the second cavity is turned off and the main RF 
is progressively tuned to nominal voltage. Before the next injection
the particles circulating around the bucket have to be collimated
at $2.5$ per mil momentum excess (fig. \ref{fig:LI-MERGING}h)
or they will hit the septum blade during the next injection.

These simulations have been done with all FP6 and FP7 ions together
with capture efficiency studies. Capture efficiency is defined as
quotient between not-collimated (e.g. running or decayed) particles
and injected particles at the end of each SPS cycle. It was possible
to reach an efficiency of slightly below 90\% in all cases (table
\ref{tab:ION-ACCUM-RESULT}). However the results for $^{8}$Li
and $^{8}$B are very preliminary since no data on longitudinal emittances
was available at the time of the simulations.

\begin{table}[htbp]
\begin{tabular}{r|cccc}
                                    & $^{6}$He & $^{18}$Ne & $^{8}$Li & $^{8}$B     \\
\hline
$t_{1/2}$ at rest [ms]               & 807     &      1872 &         838 & 770            \\
SPS cycle time [s]                  & 6.0     &       3.6 &         4.8 & 3.6            \\
Source rate [10$^{13}$/s]           & 2       & 2.09      & 9           & 9              \\
Inject./Bunch [10$^{11}$]           & 4.87    &      2.35 &       21.50 & 8.43           \\
Capture Efficiency                  & 88.8\%  &    87.8\% &      88.2\% & 89.0\%         \\
Acc./Bunch [10$^{11}$]             & 40      &        31 &         173 & 74             \\
Acc. in DR [10$^{13}$]             & 8       & 6.20      & 35          & 13.6           \\
$\nu$-rate [10$^{18}$/year]       & 2.4     & 0.92      & 10.2        & 4.34           \\
Nom. $\nu$-rate [10$^{18}$]       & 2.9     & 1.1      & 14.5        & 5.5           \\
$\nu$-rate ratio                   & 0.828   & 0.836     & 0.705       & 0.788           \\
\end{tabular}
\caption{DR parameters and simulation
results for all isotopes. Nominal $\nu$-rates for $^{8}$Li and $^{8}$B are 5 times bigger 
than for $^{6}$He and $^{18}$Ne respectively \cite{FernandezMartinez}.
Results for Lithium and Boron are preliminary. (Acc. stands for accumulated.)
}
\label{tab:ION-ACCUM-RESULT}
\end{table}

\begin{figure}[ht]
$\begin{array}{c@{\hspace{0.1in}}c}
\includegraphics[angle=0, scale= .15]{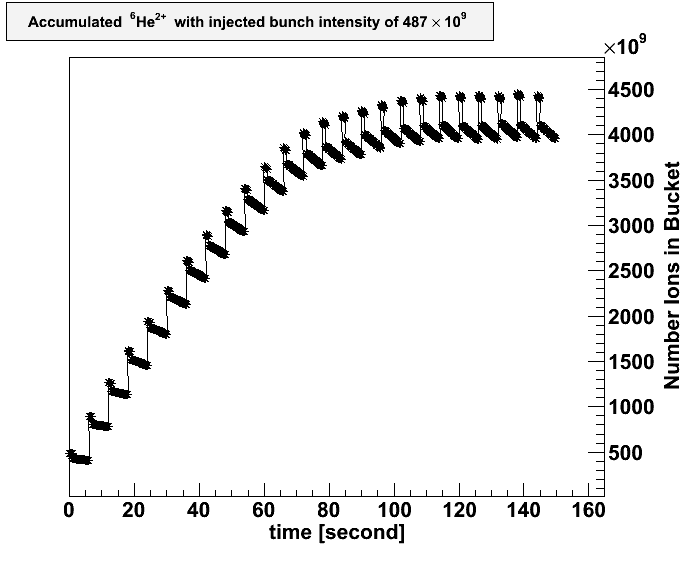}   &  \includegraphics[angle=0, scale= .15]{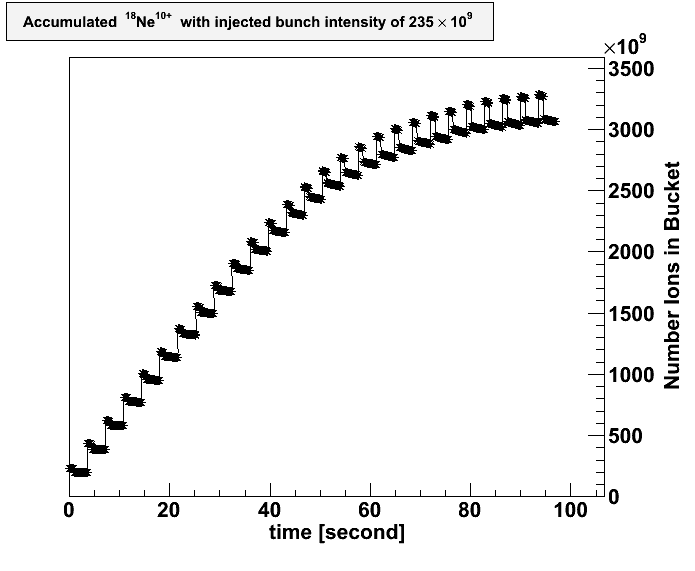} \\ [-2.7cm]
\multicolumn{1}{l}{\mbox{\bf \ \  \ \ \ \ \   (a)}}                                &  \multicolumn{1}{l}{\mbox{\bf \ \  \ \ \ \ \    (b)}} \\ [+2.3cm]
\includegraphics[angle=0, scale= .15]{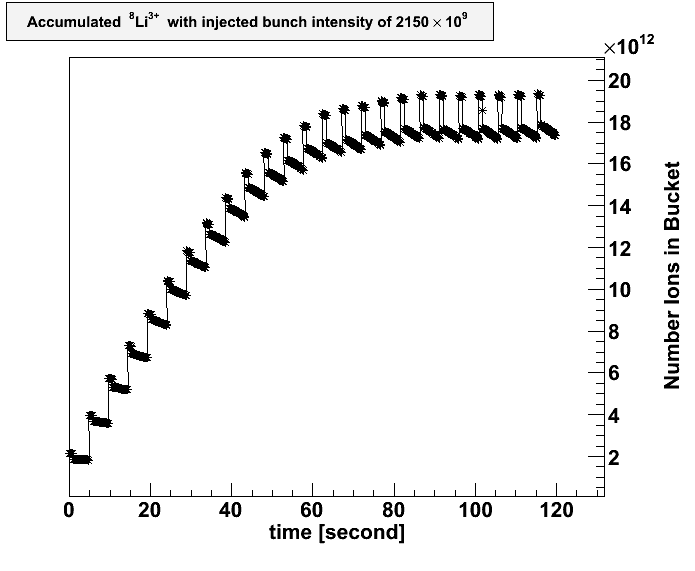}   &  \includegraphics[angle=0, scale= .15]{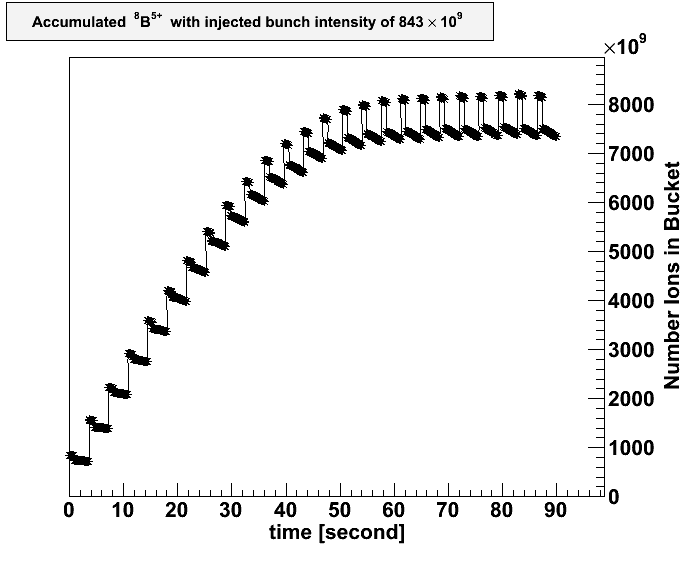} \\ [-2.7cm]
\multicolumn{1}{l}{\mbox{\bf \ \  \ \ \ \ \   (c)}}                                &  \multicolumn{1}{l}{\mbox{\bf \ \  \ \ \ \ \    (d)}} \\ [+2.3cm]
\end{array}$
\caption{Accumulation of (a) $^{6}$He, (b) $^{18}$Ne, (c) $^{8}$Li and (d) $^{8}$B ions per bunch.}
\label{fig:ACCUMULATION}
\end{figure}

During the merging process the longitudinal emittance of the stored
beam is increased due to RF gymnastics. 
The collimation at $2.5$ per mil momentum excess will therefore limit
the bunch size after a certain number of injections. 
Together with radioactive decay (also included in the BBPhase simulation)
the collimation will therefore cause the accumulated number of ions in 
the bunch to saturate. 
We obtain the number of particles after which a bunch is saturated
by repeatedly applying the merging program with the appropriate cycle
times for each ion (table \ref{tab:ION-ACCUM-RESULT}). 
The amount of remaining particles per bunch is plotted
every $4000$ turns over a period of $25$ injections ($27$
for $^{18}$Ne due to later saturation of Neon) in figure \ref{fig:ACCUMULATION}.
Due to uncertainties in the ion production (R\&D is ongoing) the amount of 
injected ions per cycle is an estimated value for all isotopes.

At saturation the size of one stored bunch is about 2 meters for all ions. 
The duty cycle of the DR ( = suppression factor of the experiment), assuming 20
injected bunches, is then $DC=20\cdot2m/6911m=0.58\%$.
This holds for all tested ions which is in agreement with the results
in \cite{CHANCE} and extends them to $^{8}$Li and $^{8}$B.
From the number of accumulated ions and 20 injected bunches
we can calculate the annual neutrino flux via a Mathematica notebook
\cite{WILDNER-MATHEMATICA}. It turns out that the amount of ions
stored at saturation ($3.1\cdot10^{12}$Ne per bunch and $4\cdot10^{12}$He
per bunch) corresponds to only $84\%$ ($83\%$) of the nominal (anti)
neutrino flux \cite{FP6-FINAL} 
(see table \ref{tab:ION-ACCUM-RESULT}).
In the case of $^{8}$Li and $^{8}$B sensitivity plots shows that nominal 
neutrino rate needed is 5 times larger than for $^{6}$He and $^{18}$Ne \cite{FernandezMartinez}.
The (anti) neutrino flux achieved from
the number of ions stored in the DR at saturation corresponds to $71\%$
($^{8}$Li) and $79\%$ ($^{8}$B) of the nominal fluxes.

We are however well below the requirements of the duty cycle which
is at maximum $1\%$.


\section{Conclusions}

It was possible to adapt the radio frequency bunch merging scheme
to the 2D program BBPhase and recreate the capture efficiencies already calculated
 in case of FP6 ions \cite{CHANCE}. These results could also
be extended to $^{8}$Li and $^{8}$B using preliminary values for
emittances and ion production showing the feasibility of bunch merging
for these isotopes. Moreover the duty factor considering 20 accumulated
bunches in the Decay Ring was found to be 0.58\% which also agrees
with the previous theoretical results and is well below the upper
limit of $1\%$. In light of these results the Bunch Merging procedure
is at present the preferred method to create ion bunches with a sufficiently
low duty cycle inside the Decay Ring.

However the amount of ions that could be stored in one bunch at saturation
was found to be lower than needed for the production of nominal (anti)
neutrino flux in the Helium and Neon case. For the Lithium and Boron
case further R\&D is needed in terms of ion production and transportation.
The simulations will then have to be repeated with new values for
longitudinal emittances and injected bunch intensities.

\begin{theacknowledgments}
We acknowledge the financial support of the European Community under the
European Commission Framework Programme 7 Design Study: EUROnu, Project
Number 212372. The EC is not liable for any use that may be made of the
information contained herein.
\end{theacknowledgments}


\bibliographystyle{aipproc} 
\bibliography{References}

\begin{thebibliography}{10}
\expandafter\ifx\csname natexlab\endcsname\relax\def\natexlab#1{#1}\fi
\providecommand{\enquote}[1]{``#1''}
\expandafter\ifx\csname url\endcsname\relax
  \def\url#1{\texttt{#1}}\fi
\expandafter\ifx\csname urlprefix\endcsname\relax\def\urlprefix{URL }\fi
\providecommand{\eprint}[2][]{\url{#2}}

\bibitem[FP7(2007)]{FP7-EURONU}
  (2007), \eprint{FP7-INFRASTRUCTURES-2007-1}.

\bibitem[Zucchelli(2002)]{ZUCCHELLI-BETABEAM}
P.~Zucchelli, \emph{Phys. Lett.} \textbf{B532}, 166--172,  (2002).

\bibitem[{EC}(????{\natexlab{a}})]{FP6-HOMEPAGE}
{EC} \eprint{http://cordis.europa.eu/fp6}.

\bibitem[{EC}(????{\natexlab{b}})]{FP7-HOMEPAGE}
{EC} \eprint{http://cordis.europa.eu/fp7}.

\bibitem[Hansen et~al.(2010)]{HANSEN}
C.~Hansen, E.~Wildner, and E.~F. Martinez, \emph{AIP Conference Proceedings}
  \textbf{1222}, 455--458,  (2010),
  \urlprefix\url{http://link.aip.org/link/?APC/1222/455/1}.

\bibitem[{Christian Hansen, Daniel Heinrich}(????)]{SVN}
{Christian Hansen, Daniel Heinrich}, {BBPhase},
  \eprint{http://svnweb.cern.ch/world/wsvn/bbphase/}.

\bibitem[Chancé et~al.(????)]{CHANCE}
A.~Chancé, et~al. 12-25-2009-0019.

\bibitem[Fernandez-Martinez(2010)]{FernandezMartinez}
E.~Fernandez-Martinez, \emph{Nucl. Phys.} \textbf{B833}, 96--107,  (2010),
  \eprint{0912.3804}.

\bibitem[Wildner et~al.(????)]{WILDNER-MATHEMATICA}
E.~Wildner, et~al. \eprint{CERN-AB-2007-015}.

\bibitem[{Beta Beam Study Group}(2009)]{FP6-FINAL}
{Beta Beam Study Group}  (2009),
  \urlprefix\url{http://beta-beam.web.cern.ch/beta-beam/task/Final\%20report/F%
inalreport.asp}.

\end{thebibliography}

\end{document}